\documentclass[onecolumn]{IEEEtran} 
\usepackage{setspace}
\usepackage{cite}
\usepackage{hyperref}
\ifCLASSINFOpdf
\usepackage[pdftex]{graphicx}
\else
   \usepackage[dvips]{graphicx}
\fi
\usepackage[cmex10]{amsmath}
\usepackage{algorithm}
\usepackage{algorithmic}
\usepackage{float}
\newfloat{algorithm}{t}{lop}
\usepackage{array}
\usepackage{booktabs}
\usepackage{url}
\usepackage{color}
\usepackage{multirow}
\usepackage{pdflscape}
\usepackage{afterpage}
\usepackage{amsmath,amssymb,amsfonts,amsthm}

\ifCLASSOPTIONcompsoc
  \usepackage[caption=false,font=normalsize,labelfont=sf,textfont=sf]{subfig}
\else
  \usepackage[caption=false,font=footnotesize]{subfig}
\fi
\usepackage{dblfloatfix}
\newcount\Comments  
\usepackage{color}
\definecolor{darkgreen}{rgb}{0,0.5,0}
\definecolor{purple}{rgb}{1,0,1}
\newcommand{\kibitz}[2]{\ifnum\Comments=0\textcolor{#1}{#2}\fi}

\newcommand{\R}{\mathrm{I\kern-0.21emR}}
\newcommand{\N}{\mathrm{I\kern-0.21emN}}

\renewcommand{\leq}{\leqslant}

\newtheorem{definition}{Definition}
 \newtheorem{lemma}{Lemma}

\begin{document}
\title{Are commercially implemented adaptive cruise control systems string stable?}
\author{George Gunter, Derek Gloudemans, Raphael E. Stern, Sean McQuade, Rahul Bhadani, Matt Bunting, Maria Laura Delle Monache, \IEEEmembership{Member}, Roman Lysecky,~\IEEEmembership{Senior Member}, Benjamin Seibold, Jonathan Sprinkle~\IEEEmembership{Senior Member}, Benedetto Piccoli,~\IEEEmembership{Senior Member}, Daniel B. Work, \IEEEmembership{Member}
\thanks{This material is based upon work supported by the National Science Foundation under Grants No. CNS-1446715 (B.P.), CNS-1446690 (B.S.), CNS-1446435 (R.L.), CNS-1446702 (D.W.) and OISE-1743772 (G.G.). This work was supported by the Inria Associated team "MEMENTO". M.L.D.M. was supported by the IDEX-IRS 2018 project "MAVIT".}
\thanks{G. Gunter is with the Department of Civil and Environmental Engineering at the University of Illinois at Urbana-Champaign, Urbana, IL 61801 USA and with the Institute for Software Integrated Systems, Vanderbilt University, Nashville, TN 37212 USA  (e-mail: gunter1@illinois.edu).}
\thanks{S. McQuade and B. Piccoli are with the department of Mathematical Sciences at Rutgers University, Camden, NJ 08102, USA (e-mail: \{stm106, piccoli\}@camden.rutgers.edu). }
\thanks{R. Bhadani, M. Bunting, R. Lysecky, and J. Sprinkle are with the Department of Electrical and Computer Engineering, University of Arizona, Tucson, AZ 85719, USA (e-mail: \{rahulbhadani, mosfet, rlysecky, sprinkjm\}@email.arizona.edu).}
\thanks{M. L. Delle Monache is with the NeCS team at Univ. Grenoble Alpes, Inria, CNRS, Grenoble INP, GIPSA-Lab, 38000 Grenoble, France, (e-mail:ml.dellemonache@inria.fr).}
\thanks{B. Seibold is with the Department of Mathematics, Temple University, Philadelphia, PA 19122 (e-mail: seibold@temple.edu).}
\thanks{D. Gloudemans, R. Stern, and D. B. Work are with the Department of Civil and Environmental Engineering and the Institute for Software Integrated Systems, Vanderbilt University, Nashville, TN 37212 USA, (e-mail: \{derek.gloudemans, raphael.stern, dan.work\}@vanderbilt.edu).}}

\maketitle

\begin{abstract}
In this article, we assess the string stability of seven 2018 model year \textit{adaptive cruise control} (ACC) equipped vehicles that are widely available in the US market. Seven distinct vehicle models from two different vehicle makes are analyzed using data collected from more than 1,200 miles of driving in car-following experiments with ACC engaged by the follower vehicle. The resulting dataset is used to identify the parameters of a linear second order delay differential equation model that approximates the behavior of the black box ACC systems. The string stability of the data-fitted model associated with each vehicle is assessed, and the main finding is that all seven vehicle models have string unstable ACC systems. For one commonly available vehicle model that offers ACC as a standard feature on all trim levels, we validate the string stability finding with a multi-vehicle platoon experiment in which all vehicles are the same year, make, and model. In this test, an initial disturbance of 6 mph is amplified to a 25 mph disturbance, at which point the last vehicle in the platoon is observed to disengage the ACC. The data collected in the driving experiments is made available, representing the largest publicly available comparative driving dataset on ACC equipped vehicles.
\end{abstract}

\section{Introduction}
\label{sec:introduction}
Adaptive cruise control systems are now widely available as a standard or optional feature on many of the best-selling cars in the US and around the world. These vehicles represent the first wave of \textit{Society of Automotive Engineers} (SAE) level 1 automated vehicle systems beginning to appear in the traffic flow. Changing the car following dynamics of a small fraction of vehicles in the traffic flow can fundamentally change the emergent properties of the flow, as experimentally demonstrated by Stern et al.~\cite{stern2017dissipation}, where a single autonomous vehicle was used to stabilize the traffic flow and eliminate stop-and-go waves on a ring track. More broadly, the interest in determining the potential benefits of automated driving systems on the traffic flow has been an ongoing research focus for the vehicular control and traffic engineering communities~\cite{darbha1999,bose2003analysis,davis2004effect,kesting2010enhanced,shladover2012impacts,van2006impact,delis2016simulation,talebpour2016influence,dollar2018efficient}. 

Despite many positive theoretical and simulation driven findings about the benefit of \textit{adaptive cruise control} (ACC) systems on traffic flow throughput and flow stability to date, it is not known if these benefits are achievable with current vehicles that are in the market today. The main question this research article addresses is whether the currently available commercial ACC systems amplify or dissipate small disturbances through a platoon of vehicles. In a string stable platoon of vehicles, small perturbations will be dissipated as they propagate from one vehicle to another, while in a string unstable platoon small perturbations from equilibrium may amplify as they propagate through the platoon. String unstable adaptive cruise control systems can lead to the presence of phantom traffic jams~\cite{FlynnKasimovNaveRosalesSeibold2009} that seemingly appear without cause, similar to the ones caused by human drivers~\cite{Sugiyamaetal2008}.

Consequently, beyond safety and rider comfort~\cite{xiao2010comprehensive}, a key challenge in automated driving systems is to design control laws in which the vehicle platoon remains string stable, a question for which significant theoretical and practical progress has been made~\cite{levine1966optimal,ioannou1993intelligent,shladover1995review,swaroop1996string, rajamani1998design,liang1999optimal, alam2015heavy,orosz2005bifurcations,besselink2017string,monteil2018l2}. For example, it is well known that constant spacing policies lead to string unstable platoons of vehicles~\cite{seiler2004disturbance,jovanovic2005ill,middleton2010string}. Alternatively, by relaxing the requirement of rigid platoon formations, constant time-headway policies can achieve string stability~\cite{ioannou1993autonomous,liang1999optimal} and serve as a basis for ACC implementations. 

One possible way to improve constant spacing-based platoons of vehicles is to enhance each vehicle with connectivity to other vehicles in the platoon. It has been shown that if a platoon of vehicles is connected and automated, then it is possible to form dense platoons of vehicles which leave very small gaps. Recent work has experimentally demonstrated the benefits of connected adaptive cruise control systems to achieve string stability~\cite{ploeg2011design,milanes2014cooperative,naus2010string}, even when human driven vehicles are also present~\cite{jin2018connected}. However, connectivity of this form is not yet available on commercially available vehicles.  

The main contribution of this work is therefore the field testing of seven commercially available ACC systems to answer the question raised in the title of the article: \textit{are commercially implemented adaptive cruise control systems string stable}? We find, across two makes and seven vehicle models, all available in 2018, that the answer is: ``no.'' All ACC systems tested in this work are found to be string unstable.
        
The experimental setup in our work is inspired by the work of Bareket et al.~\cite{bareket2003methodology} and Milan\'es and Shladover~\cite{milanes2014modeling}. We require a lead vehicle to drive at specified speed profiles, and a test vehicle follows the lead vehicle with its ACC engaged. The present article has a different focus than~\cite{milanes2014modeling}, which illustrated how connected ACC systems can be designed to be string stable. It also builds on the work~\cite{gunter2019}, which found a single luxury electric vehicle ACC system to be string unstable. In the present article, we expand on~\cite{gunter2019} by \textit{i}) testing seven commercial ACC systems and providing the experimental data, bringing the total number of commercial ACC systems tested to nine (in addition to the commercial systems tested in~\cite{milanes2014modeling} and~\cite{gunter2019}); \textit{ii}) showing that all of the tested systems are string unstable, which is a negative result for phantom traffic jam prevention; \textit{iii}) confirming the consequences of the string instability in simulation and also with with a large, eight vehicle platoon test (one leader followed by seven ACC engaged vehicles) in which a small 6 mph disturbance is amplified to a 25 mph disturbance before the last vehicle automatically disengaged the ACC system. This is in contrast to platoon simulations reported in~\cite{gunter2019}, which indicated that the string unstable ACC system may still reduce some disturbances for moderate sized platoons of up to 15 vehicles. 

The remainder of this article is structured as follows. In Section~\ref{sec:dynamics_model} we describe an analytical model which is used to approximate the behavior of the car following dynamics of the test vehicle operating with ACC engaged. We establish the notation and describe the theoretical background for the string stability analysis, as well as describe the methods used to fit the model to the experimental data. In Section~\ref{sec:exper_method} we describe in detail the data collection experiments conducted on each of the seven ACC vehicles, and also explain the methods used to set up a eight vehicle platoon test with one lead vehicle and seven ACC following vehicles of the same make and model to validate the string stability findings for one of the vehicle models. In Section~\ref{sec:results} we present the results obtained for the model calibration and the stability analysis, finding all vehicles are string unstable. Via simulation of platoons made up of identical vehicles using the calibrated models, we illustrate the wide range of behavior of vehicle platoons with respect to the same disturbance. We also confirm via a platoon experiment with real ACC vehicles that a small initial disturbance can grow large enough to cause the ACC system to disengage further back in the platoon.

\section{Adaptive cruise control dynamics, stability and data fitting}\label{sec:dynamics_model}
In this section, a method is presented for determining the string stability of ACC vehicles from data. We propose a car following model to approximate the driving behavior of the ACC vehicle with parameters that can be calibrated from data. We then analyze the string stability of the calibrated model, which offers a proxy for analyzing the stability of the black box code being executed on the vehicles themselves.

\subsection{Model definition}
Car following models are regularly used in the traffic engineering community to approximate the behavior of human drivers, and also to approximate the behavior of automated vehicles~\cite{talebpour2016influence,kesting2010enhanced,shladover2012impacts}. The benefit of this modeling choice is that the resulting differential equations models are straightforward to develop and are amenable to analysis. Many different car following models have been proposed in the literature, such as the intelligent driver model \cite{treiber2000congested, treiber2006delays} and the Gipps model \cite{gipps1981behavioural}. In this work, the ACC vehicle dynamics and driving behavior of a platoon of $N$ vehicles indexed by $i$ are modeled using a variation on a common \textit{optimal velocity} micro-model~\cite{BandoHesebeNakayama1995} with a \textit{relative velocity} term (OVRV):

\begin{equation}
    \label{eq:OVRV}
    \left\lbrace
    \begin{array}{ll}
         \dot{s}_i(t)&= \Delta v_i(t), \qquad i=1,\ldots, N  \\
         \dot{v}_i(t)&= k_1 \lbrack V(s_i(t))-v_i(t)\rbrack+k_2 \lbrack\Delta v_i(t) \rbrack.
    \end{array}
    \right.
\end{equation}
Here the acceleration of vehicle $i$, denoted $\dot{v}_i$, is adjusted proportional to the difference between the current velocity $v_i$ and the desired velocity according to the optimal velocity function $V(s_i)$ for space gap $s_i$, and the speed difference, $\Delta v_i = v_{i-1}-v_i$, between the vehicle $i$ and a vehicle  $i-1$ in front.  The parameters $k_1$ and $k_2$ are parameters representing the gains on the two terms. The velocity $v_0$ is the lead vehicle speed  that we assume to have fixed dynamics with a piecewise constant speed. 

An OVRV model commonly used to model ACC systems is~\cite{liang1999optimal,milanes2014cooperative}: 
\begin{equation}
    \label{eq:CTH-FL}
    \left\lbrace
    \begin{array}{ll}
         \dot{s}_i(t)&= \Delta v_i(t), \qquad i=1,\ldots, N  \\
         \dot{v}_i(t)&= k_1 \lbrack s_i(t)- \eta - t_{h} v_i(t)\rbrack + k_2 \lbrack \Delta v_i(t)\rbrack.
    \end{array}
    \right.
\end{equation}
Here $t_{h}$ represents a desired effective time gap that the ACC seeks to maintain, and $\eta$ is the jam space gap (i.e., the desired space gap when the vehicles are at rest). The model obtains a steady state when all vehicles have the same sped (i.e., $\Delta v_i=0$), and each vehicle has an effective space gap equal to the desired effective time gap to the car in front of it (i.e., $t_{h}=(s_i-\eta) / v_i$). 

The model is further modified to allow for a time delay $\tau$, which accounts for systematic delay in system sensors. This results in the delay differential equation:
\begin{equation}
    \label{eq:acc_model}
    \left\lbrace
    \begin{array}{ll}
         \dot{s}_i(t)&= v_{i-1}(t)-v_i(t), \qquad i=1,\ldots, N  \\
         \dot{v}_i(t)& = k_1 \lbrack s_i(t-\tau)- \eta - t_{h} v_i(t)\rbrack  \\&
         ~~~~~~~+ k_2 \lbrack v_{i-1}(t-\tau)-v_i(t)\rbrack.
    \end{array}
    \right.
\end{equation}

In \eqref{eq:acc_model}, each vehicle is assumed to measure the space gap and the leader velocity with delay (e.g., due to sampling rates, and data processing). We assume the delay on the sensors measuring properties of the lead vehicle are large relative to any delays of sensors measuring the velocity of the vehicle itself. From that, the vehicle accelerates or decelerates to match the desired space gap and the leader velocity. This DDE is used throughout the remaining of the paper as the model to describe driving behavior and dynamical responses of different ACC vehicles.

\subsection{Model stability analysis}\label{sec:stability}
In this section, we describe how to analyze the string stability of platoons of vehicles with ACC engaged, assuming each vehicle in the platoon is described by the dynamics~\eqref{eq:acc_model}. We consider a platoon of $N$ vehicles on a closed ring in which vehicle $i$ follows vehicle $i-1$ (and vehicle $1$ follows vehicle $N$), based on how we have spacing and the relative velocity defined. The use of the ring for mathematical convenience is justified since, as we describe below, a platoon of vehicles which exhibits unstable dynamics on the ring is string unstable on a straight road.

First, we investigate the asymptotic stability of \eqref{eq:acc_model} at equilibrium on a ring road of length $L$. On the ring,~\eqref{eq:acc_model} has a unique equilibrium point given by

\begin{equation}
    \label{eq:equil_point}
    \begin{array}{ll}
         &s_i(t)\equiv s^*=\frac{L}{N}, \qquad v_i(t)\equiv v^*=\frac{s^*-\eta}{t_{h}}\\
    \end{array}
\end{equation}
This allows us to rewrite the system~\eqref{eq:acc_model} in a new reference frame looking at the spacing and velocity perturbations: $\tilde{\xi}_i(t) =[\tilde{s}_i(t), \tilde{v}_i(t)] = \xi_i(t)-\xi_i^*$,  with $\xi_i(t) = [s_i(t),v_i(t)]$ and $\xi^*_i(t) = [s^*,v^*]$ :

\begin{equation}\label{eq:equilibrium_system}
\dot{\tilde{\xi}}_i(t) = 
\begin{pmatrix}
\tilde{v}_{i-1}(t)-\tilde{v}_i(t)\\
k_1[(\tilde{s}_i(t-\tau)-\eta-t_{h} \tilde{v}_i(t)] \\
         ~~~~~~~+ k_2[\tilde{v}_{i-1}(t-\tau)-\tilde{v}_i(t)].
\end{pmatrix}
\end{equation}

Following the work of \cite{jin2014dynamics, zhang2016motif}, we look for plant and head-to-tail string stability to characterize the system performances. They are defined as follows.
 \begin{definition}
 The system is said to be plant stable if the equilibrium of system \eqref{eq:acc_model} is asymptotically stable when there are no external disturbances.
 \end{definition}
 \begin{definition}
 When disturbances are imposed on the head vehicle, the system is said to be string stable if the disturbances are attenuated when reaching the tail vehicle. 
 \end{definition}
 
We consider the velocity perturbation of the vehicle $\tilde{v}_{i}$ of the head vehicle as the input and the velocity perturbation of $\tilde{v}_{i+1}$ of the tail as the output. 
Taking the Laplace transform of system \eqref{eq:equilibrium_system} with zero initial condition we obtain the head-to-tail transfer-function
\begin{equation}
     \label{eq:transform}
     \Gamma(z) = \dfrac{\tilde{V}_{i+1}(z)}{\tilde{V}_i(z)}
\end{equation}
where $\tilde{V}_{i+1}(z)$ and $\tilde{V}_i(z)$ are the Laplace transform of $\tilde{v}_{i+i}$ and $\tilde{v}_i$.
\begin{lemma}
System \eqref{eq:equilibrium_system} is plant stable if and only if all solutions of the characteristic equation $\Gamma(z)=0$ are located in the left half complex plane, \cite{wilson2011car}.

\end{lemma}
\begin{lemma}
If \eqref{eq:equilibrium_system} is asymptotically stable then the equivalent system on an open road might be string stable (unstable) \cite{monteil2018l2}.
If the closed loop system \eqref{eq:equilibrium_system} is asymptotically unstable, then the system with the same controls on a straight road cannot be string stable.
\end{lemma}

\begin{figure}
    \centering
    \includegraphics[trim=60 30 60 0,clip,width = 0.7 \columnwidth]{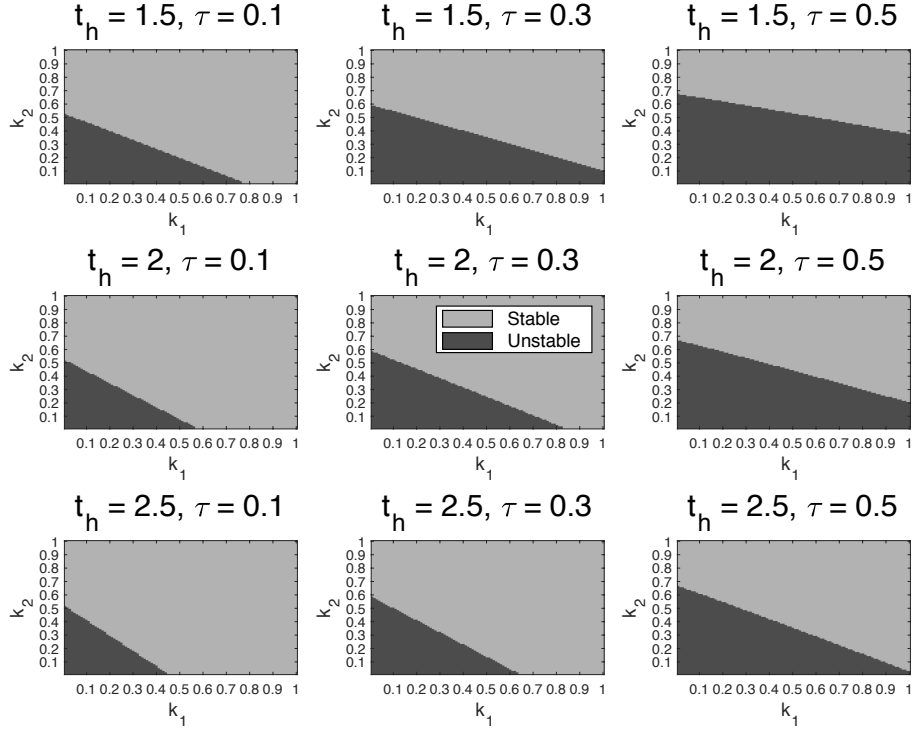}
    \caption{String stability diagrams in the $(k_1-k_2)$-plane for different values of $\tau$ and $t_h$ with $\eta = 10$.}
    \label{fig:traffic_stability}
\end{figure}

Using the continuation package DDE-BIFTOOL, \cite{engelborghs2001dde}, we analyze the string stability of model \eqref{eq:equilibrium_system} on a closed loop road in the space parameter $(k_1, k_2)$, fixing values of the delay $\tau$, the headway $t_h$, and the jam space gap $\eta$. 
DDE-BIFTOOL is a set of routines for Matlab which, among other things, provides a tool to perform numerical bifurcation analysis of steady state and periodic solutions for differential equations with constant delays.
It can be seen in Figure~\ref{fig:traffic_stability} that there are choices of parameters for which system \eqref{eq:equilibrium_system} on a closed loop road is not asymptotically stable, which implies that the equivalent system on an open road is string unstable. For example, the model of the form~\eqref{eq:acc_model} with $t_h = 1.5$ s, $\tau = 0.1$ s, $k_1 = 0.2$ 1/s$^2$, and $k_2 = 0.2$ 1/s and $\eta = 10$ m, this model represents a string unstable model, as seen in Figure~\ref{fig:traffic_stability}. Consequently, vehicles following~\eqref{eq:acc_model} under these parameters will also amplify disturbances on the line.

\subsection{Model calibration}\label{sec:system_id}

To study the driving behavior and dynamics of ACC equipped vehicles, the parameters $k_1$, $k_2$, $t_{h},\tau$ and $\eta$ in \eqref{eq:acc_model} must be calibrated to best reproduce experimental data. To determine the model parameters, an error metric is used to compare the performance of the model to the observed data. Here we consider the \textit{mean square error} (MSE) on the velocity as the performance measure:
\begin{equation}
\label{eq:calibrationModel}
\mathrm{MSE} = \dfrac{1}{T}\int_0^T\left(v(t) - v^\text{m}(t)\right)^2 dt,
\end{equation}
where $v^\text{m}(t)$ is the measured velocity of the follower ACC vehicle, $v$ is the simulated velocity of the same vehicle from the model, and $T$ is the duration of the experiment.

With the performance criterion defined, the optimal parameters can be determined by solving the following optimization problem:
\begin{equation}\label{eq:minCTH}
\begin{array}{rl}
\underset{s,v,k_{1},k_{2},t_{h},\tau,\eta}{{\text{minimize}}}: & \dfrac{1}{T}\int_\tau^T\left(v(t) - v^\text{m}(t)\right)^2dt\\
\text{subject to:}& \dot{s}(t)= v^\text{m}_\ell(t)-v(t), t\in[\tau,T]\\
 & \dot{v}(t)=k_1 \lbrack s(t-\tau) - \eta - t_{h} v(t)\rbrack \\&
 ~~~~~~~+ k_2 \lbrack v^\text{m}_\ell (t-\tau)-v(t)\rbrack, t\in[\tau,T] \\
 & s(t)=s^\text{m}(t),\forall t\in[0,\tau]\\
 & v(t)=v^\text{m}(t), \forall t\in[0,\tau]\\
 & k_1^\text{l} \leq k_1 \leq k_1^\text{u}\\
 & k_2^\text{l} \leq k_2 \leq k_2^\text{u}\\
 & \tau^\text{l} \leq \tau \leq \tau^\text{u}\\
 & t_h^\text{l} \leq t_h\leq t_h^\text{u}\\
 & \eta^\text{l} \leq \eta \leq \eta^\text{u},
 \end{array}
\end{equation}
where velocity and space gap of the follower are denoted $v$ and $s$ respectively. Here $s^\text{m}(t)$ and $v^\text{m}(t)$ denote the measured space gap and velocity from the experimental data, and $k_1$, $k_2$, $t_{h},\tau$ and $\eta$ are the above mentioned model parameters, which are decision variables for the problem.  The measured leader velocity is denoted  $v^\text{m}_\ell$, respectively. The problem is constrained with lower (denoted with a superscript $\text{l}$) and upper (denoted with a superscript $\text{u}$) bounds on the model parameters.

\section{Experimental methods}\label{sec:exper_method}
In this section, we briefly describe the vehicle fleet tested as well as the experimental design. We describe two-vehicle car following tests in which a lead vehicle drives a given velocity profile with the test ACC vehicle following behind. These tests are used to calibrate and test the quality of fit of the assumed ACC model~\eqref{eq:acc_model}. We also describe the setup of an eight vehicle platoon experiment in which seven identical ACC vehicles follow a lead vehicle that creates a velocity slow down event. This test is used to validate the string stability findings for the most common vehicle tested.  

In all experiments, a lead vehicle is used to drive with a pre-specified speed profile, and follower vehicles drive with ACC engaged behind the lead vehicle forming a single lane platoon. Thus, longitudinal control of the follower vehicle(s) is achieved by the ACC system. In total, over 1,200 miles of driving are recorded throughout the experiments. All data is openly available for public use online~\cite{data}.

\subsection{Vehicle fleet}
The vehicles tested in this experiment are all widely available, 2018 model year vehicles. Key features of the test vehicles are summarized in Table~\ref{tab:vehicle_summary}. Vehicles are from one of two manufacturers, denoted Make~1 and Make~2, in Table~\ref{tab:vehicle_summary}. Note that six of the seven vehicles tested are traditional internal combustion engine vehicles, while one vehicle is a hybrid electric vehicle. Also note that the vehicles from manufacturer~1 have a minimum ACC operating speed of 25 mph, while the vehicles from manufacturer~2 are capable of coming to a complete stop under ACC. For consistency across vehicle makes, all testing is conducted above the minimum cutoff speed of Make~1.

\begin{table}
    \centering
    \begin{tabular}{c|c c c c}
    \toprule
        Vehicle & Make & Style & Engine & Min. ACC\\
                &       &       &       & speed (mph)\\
        \midrule
         A & 1 & Full-size sedan & Combustion & 25\\
         B & 1 & Compact sedan & Combustion & 25\\
         C & 1 & Compact hatchback & Hybrid & 25\\
         D & 1 & Compact SUV & Combustion & 25\\
         E & 2 & Compact SUV & Combustion & 0\\
         F & 2 & Mid-size SUV & Combustion & 0\\
         G & 2 & Full-size SUV & Combustion & 0\\
         \bottomrule
    \end{tabular}
    \vspace{0.1cm}
    \caption{Summary of tested vehicles.}
    \label{tab:vehicle_summary}
\end{table}

\subsection{Data collection}
Position and speed data for each vehicle were collected using high-accuracy uBlox EVK-M8T GPS receivers with the antenna affixed to a known position on each vehicle. Preliminary testing of the GPS receivers indicated that the GPS receivers have a mean position accuracy of 0.24 m and a speed accuracy of 0.002 m/s error for speed. The location of the GPS antenna on each vehicle was recorded before the each experiment to accurately calculate the space gap.

\subsection{Two-vehicle tests}\label{sec:two_veh_tests}
Four speed profiles are recorded to observe the behavior of each vehicle in the two-vehicle tests. For all tests, the vehicles begin on the track and start at a low speed with a full size sedan as the lead vehicle, and the test vehicle as the follower vehicle. For consistency and comparability, the same lead vehicle is used in each test. In each two-vehicle test, the specified lead vehicle speed was implemented by setting the lead vehicle's cruise control to the desired speed. When changing speed, the manual input button was used to adjust the cruise control set point speed of the lead vehicle to the new desired speed at defined time intervals.

The two-vehicle tests are designed to fulfill two goals: (i)~obtain steady-state data to understand each vehicle's equilibrium following behavior, and (ii)~obtain transient data to understand each vehicle's transient behavior under changing lead vehicle speeds and changing space gap. The speed profiles are as follows: 

\begin{itemize}
    \item \textbf{Oscillatory:} The oscillatory test is designed to collect transient data to understand how the ACC system behaves under non-constant headway and lead vehicle speed. Therefore, for this test, both 2.7 m/s (6 mph) and 4.5 m/s (10 mph) speed fluctuations are tested. For the first half of the test the speed is fluctuated between 24.5 m/s (55 mph) and 21.9 m/s (49 mph), with each speed being held for at least 30 seconds. For the second half of the test the speed is fluctuated between 24.5 m/s (55 mph) and 20.1 m/s (45 mph) with each speed being held for at least 30 seconds.
    
    \item \textbf{Low speed steps: } The goal of this test is to collect steady-state following behavior at a broad range of speeds. For logistical reasons, steady-state data collection is divided into high and low speed tests. Therefore, this test is designed to collect low-speed steady-following behavior. Vehicles begin at 15.6 m/s (35 mph) and maintain this speed for 60 seconds at which point the speed is increased to (17.9 m/s) 40 mph and held for 60 seconds. Next, the speed is increased to 20.1 m/s (45 mph), which is held for 60 seconds, and then increased to 22.4 m/s (50 mph) and held for 60 seconds. Finally, the speed is increased to 24.6 m/s (55 mph), which is held for 60 seconds. The same speeds are next tested in reverse order (24.6 m/s, 22.4 m/s, 20.1 m/s, 17.9 m/s, 15.6 m/s), with each being held for 60 seconds.
    
    \item \textbf{High speed steps:} This test is designed to collect steady-state following behavior at high speeds. Vehicles begin at 29.1 m/s (65 mph), which is held for at least 60 seconds, then increased to 31.3 m/s (70 mph), which is held for 60 seconds, and finally increased to 33.5 m/s (75 mph) and held for at least 60 seconds. Next the same speeds are tested in a decreasing order (33.5 m/s, 31.3 m/s, and 29.1 m/s) with each held for at least 60 seconds.
    
    \item \textbf{Speed dips:} The goal of this test is to collect following behavior data for sudden changes in lead vehicle speed. Both vehicles begin at 24.6 m/s (55 mph) and hold that speed for at least 45 seconds. For this test, four different speed dips are tested: 2.7 m/s (6 mph), 4.5 m/s (10 mph), 6.7 m/s (15 mph), and 8.9 m/s (20 mph). Each speed dip is held for 5 seconds before returning to 24.5 m/s (55 mph) for at least 45 seconds. Each speed dip is conducted twice before proceeding to the next speed dip. Additional speed dips are conducted once each speed dip has been conducted at least twice, as space permits at the test site.
\end{itemize}

Note the major difference between the oscillatory test and the speed dips is that the low speed of the lead vehicle in the oscillatory tests is held for at least 30 seconds, while the lead vehicle in the dips test begins to accelerate after only five seconds. The two-vehicle tests are conducted on a 16 km (10 mile) road on flat terrain with no sharp turns or curves. High speed tests are conducted on a 16 km (10 mile) section of straight highway with little elevation change.

\subsection{Autonomous test lead vehicle}
In the eight vehicle platoon experiment, a very precise lead vehicle braking profile is desired, hence the Cognitive and Autonomous Test Vehicle (The CAT Vehicle), seen in Figure~\ref{fig:platoon_picture}, is used to achieve this. The CAT Vehicle is a modified Ford Hybrid Escape vehicle capable of operating either autonomously or under the control of a human driver. The CAT Vehicle has an integrated TORC ByWire XGV drive-by-wire platform that utilizes the \textit{Joint Architecture for Unmanned Systems} (JAUS) protocol for communications. The drive-by-wire platform consists of multiple hardware--software subsystems control modules, a central embedded controller along with a TORC SafeStop ES-220 multilevel wireless emergency stop system design to send pause and stop commands in case of an emergency. With the closed loop drive-by-wire control, a user can command desired acceleration, speed or steering control to the CAT Vehicle for autonomous operation. All commands at the lower level use the JAUS protocol to control the CAT Vehicle.

\subsection{Eight vehicle platoon test}~\label{sec:platoon_exp}
\begin{figure}
    \centering
    \includegraphics[width = 0.7 \columnwidth]{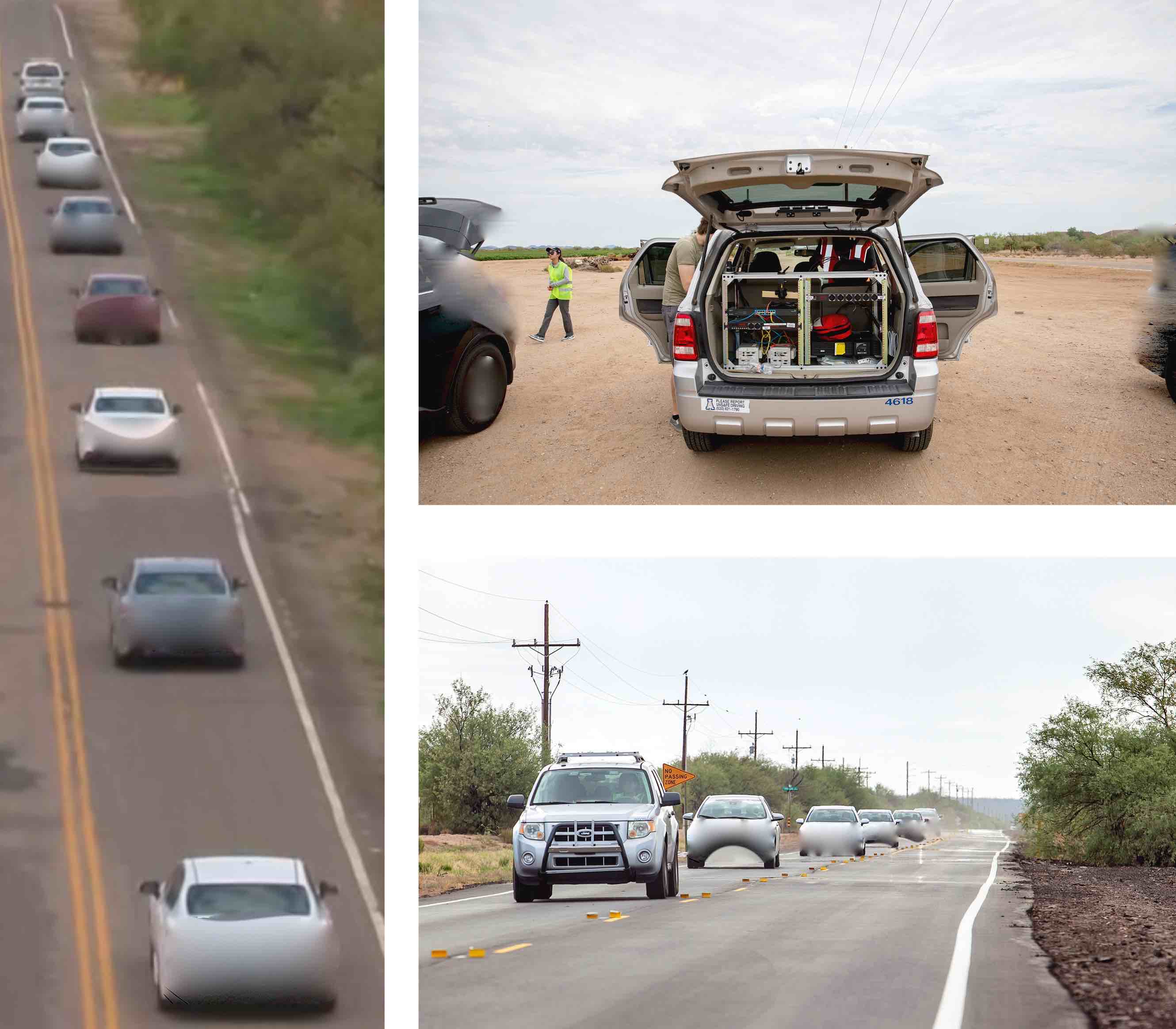}
    \caption{Photographs of the field test conducted including overhead view of platoon of CAT Vehicle followed by 7 identical vehicles (left) as well as the back of the CAT Vehicle (top right) and a ground-level view of the vehicle platoon (bottom right). Test vehicle fronts and backs blurred to remove vehicle branding.}
    \label{fig:platoon_picture}
\end{figure}

To validate the emergent traffic flow behavior of ACC vehicles at the aggregate (system) level, a platoon test is conducted where the lead vehicle executes a specific pre-specified speed profile, and seven follower vehicles drive in a single lane forming a large platoon.  An image of the vehicle platoon, captured from an overhead vantage point can be seen on the left in Figure~\ref{fig:platoon_picture}.  The CAT Vehicle is used since it is capable of consistently executing velocity commands and allows for more control over the deceleration rate of the lead vehicle in the platoon experiments.

The platoon test starts with the lead vehicle driving at 22.4 m/s (50 mph) and seven vehicles of type Vehicle~A following with the ACC engaged and the following setting at the minimum setting (i.e., the setting that allows the vehicle to follow closest to the vehicle ahead). Once all vehicles have reached steady state behavior, the lead vehicle quickly decelerates to 19.7 m/s (44 mph) and the behavior of the follower vehicles is observed. The platoon of vehicles is followed by a safety chase vehicle that keeps a larger space gap than the vehicles in the test and is used to monitor and ensure the safety of the overall experiment.

During each test, the driver of each vehicle was able to receive basic safety messages from the experiment support staff. The experiment support staff included experiment supervisor and support team, a safety supervisor and support team, and the CAT Vehicle support team responsible for overall experiment operations, safety, and management of the CAT Vehicle, respectively. Communications between the various teams and drivers was achieved via a two-way radio placed in each vehicle. For safety reasons drivers were not permitted to transmit messages while driving the vehicle, and thus were only able to receive information through the two-way radio, and not able to send information. Two-way communications were possible between the experiment, safety, and CAT Vehicle supervisors and support teams, for example to coordinate information on when to change the set point velocity. It was also used to broadcast messages to drivers to provide sufficient warning of changes in the set point speed of the lead vehicle.

\section{Results}\label{sec:results}
In this section, we first provide the results of the model calibration to the test data and determine the string stability of the seven ACC systems under minimum and maximum following settings. The consequences of the string stability findings are assessed via platoon simulations, highlighting the variability of growth rates of the vehicle disturbances. Finally, the results from the platoon experiment are used to validate the platoon simulations.

\subsection{ACC model calibration results}

\begin{table}
\centering
\begin{tabular}{c c|c c c c c c}
\toprule
Veh. & Follow & $k_1$ & $k_2$  & $t_h$  & $\tau$   & $\eta$  & Stability\\
~ & setting & [1/s$^2$] & [1/s] & [s] & [s] & [m] & ~ \\
\midrule
A & min & 0.052 & 0.338 & 0.819 & 0.948 &  8.030  & Unstable\\
A & max & 0.012 & 0.167 & 2.054 & 0.992 &  5.960  & Unstable\\
B & min & 0.052 & 0.190 & 0.725 & 0.468 &  6.849  & Unstable\\
B & max & 0.022 & 0.116 & 2.020 & 0.153 &  8.210  & Unstable\\
C & min & 0.029 & 0.269 & 0.907 & 0.368 &  10.070 & Unstable\\
C & max & 0.018 & 0.152 & 1.986 & 0.324 &  13.814 & Unstable\\
D & min & 0.051 & 0.280 & 0.544 & 0.284 &  13.400 & Unstable\\
D & max & 0.022 & 0.221 & 1.853 & 0.935 &  14.956 & Unstable\\
E & min & 0.051 & 0.165 & 1.127 & 0.419 &  5.170  & Unstable\\
E & max & 0.053 & 0.142 & 1.785 & 0.839 &  9.370  & Unstable\\
F & min & 0.071 & 0.191 & 0.696 & 0.582 &  10.090 & Unstable\\
F & max & 0.041 & 0.164 & 1.734 & 0.922 &  6.033  & Unstable\\
G & min & 0.070 & 0.253 & 0.549 & 0.993 &  14.500 & Unstable\\
G & max & 0.046 & 0.129 & 1.764 & 0.994 &  5.131  & Unstable\\
\bottomrule
\end{tabular}
\vspace{0.1cm}
\caption{Calibrated parameter values for~\eqref{eq:acc_model} for each vehicle and stability result for a two-vehicle platoon using model \eqref{eq:equilibrium_system} with learned parameters.}
\label{tab:calibration_stability_results}
\end{table}

\begin{table*}
\centering
\begin{tabular}{c c|c c c c c |c c c c c}
\toprule
~~   &  ~~ &   &    &   Speed [m/s] &   &   &    &    &   Spacing [m] &    &   \\
Veh. & Follow  & Train & Oscillatory & Low & High & Dips & Train & Oscillatory & Low & High & Dips\\
~~   & ~~  & ~~ & ~~ & speed & speed & ~~ & ~~ & ~~ & speed & speed & ~~ \\
\midrule
A & min & 0.087 & 0.201 & 0.156 & 0.100 & 0.363 & 0.505 & 1.293 & 1.120 & 1.673 & 2.067\\
A & max & 0.248 & 0.293 & 0.286 & 0.155 & 0.453 & 1.479 & 2.418 & 3.274 & 1.583 & 3.587\\
B & min & 0.342 & 0.414 & 0.285 & 0.422 & 0.509 & 4.740 & 4.860 & 2.693 & 6.157 & 6.553\\
B & max & 0.368 & 0.526 & 0.271 & 0.427 & 0.651 & 3.527 & 5.086 & 2.716 & 4.999 & 6.526\\
C & min & 0.152 & 0.409 & 0.427 & 0.214 & 0.743 & 0.826 & 2.232 & 5.224 & 2.300 & 9.532\\
C & max & 0.191 & 0.314 & 0.347 & 0.267 & 0.498 & 0.978 & 1.889 & 4.138 & 3.212 & 5.553\\
D & min & 0.228 & 0.346 & 0.260 & 0.372 & 0.433 & 2.341 & 3.569 & 2.152 & 5.326 & 4.526\\
D & max & 0.237 & 0.321 & 0.217 & 0.287 & 0.495 & 1.380 & 1.571 & 4.450 & 4.212 & 4.146\\
E & min & 0.129 & 0.302 & 0.207 & 0.294 & 0.459 & 0.803 & 1.905 & 2.796 & 1.906 & 2.611\\
E & max & 0.167 & 0.198 & 0.225 & 0.233 & 0.476 & 1.355 & 1.555 & 1.418 & 2.867 & 3.945\\
F & min & 0.206 & 0.391 & 0.229 & 0.171 & 0.756 & 1.269 & 1.991 & 1.302 & 1.640 & 5.150\\
F & max & 0.215 & 0.293 & 0.151 & 0.430 & 0.234 & 1.249 & 1.634 & 1.672 & 4.042 & 1.265\\
G & min & 0.172 & 0.278 & 0.172 & 0.211 & 0.463 & 0.944 & 1.524 & 1.879 & 1.299 & 2.353\\
G & max & 0.209 & 0.263 & 0.218 & 0.192 & 0.563 & 1.210 & 1.596 & 1.503 & 3.632 & 2.633\\
\bottomrule
\vspace{0.1cm}
\end{tabular}
\caption{Test error on all data sets collected in experiments -- trained on the first quarter of the oscillatory data only.}
\label{tab:test_errors}
\end{table*}

To develop models for each vehicle that can then be analyzed for stability, the model calibration problem outlined in Section~\ref{sec:system_id} must be solved using the available data. This is conducted by numerically solving the constrained optimization problem~\eqref{eq:minCTH} on a set of training data. The particular training data set used in this calibration is selected as the first half of the 2.7 m/s (6 mph) low-speed oscillatory data. The remaining data is left out as test data, and errors for models are displayed in Table~\ref{tab:test_errors}.

The optimization problem proposed in~\eqref{eq:minCTH} is solved using the \textit{Nonlinear Optimization with the MADS} (NOMAD) solver~\cite{AuLeTr09a, Nomad} using the \textit{Matlab} delay-differential equation solver \textit{DDE23}. The NOMAD solver uses a mesh-adaptive direct search method for numerical optimization.

The calibration problem includes upper and lower bounds on the parameters given by $k_1$, $k_2$, $t_{h}, \tau$ and $\eta$. In this work, the following bounds are used: $(k_1^\text{l},k_1^\text{u})~=~(0,1)$~1/s$^2$, $(k_2^\text{l},k_2^\text{u})~=~(0,1)$~1/s, $(t_{h}^\text{l},t_{h}^\text{u})~=~(0,3)$~s, $(\tau^\text{l},\tau^\text{u})~=~(0,1)$~s, $(\eta^\text{l}, \eta^\text{u})~=~(5,15)$~m. These limits are selected to reduce the search space of the NOMAD solver by eliminating parameters that are likely to lead to unrealistic car following behavior.

\begin{figure}
    \centering
    \includegraphics[trim=80 0 80 0,clip,width = 0.7\columnwidth]{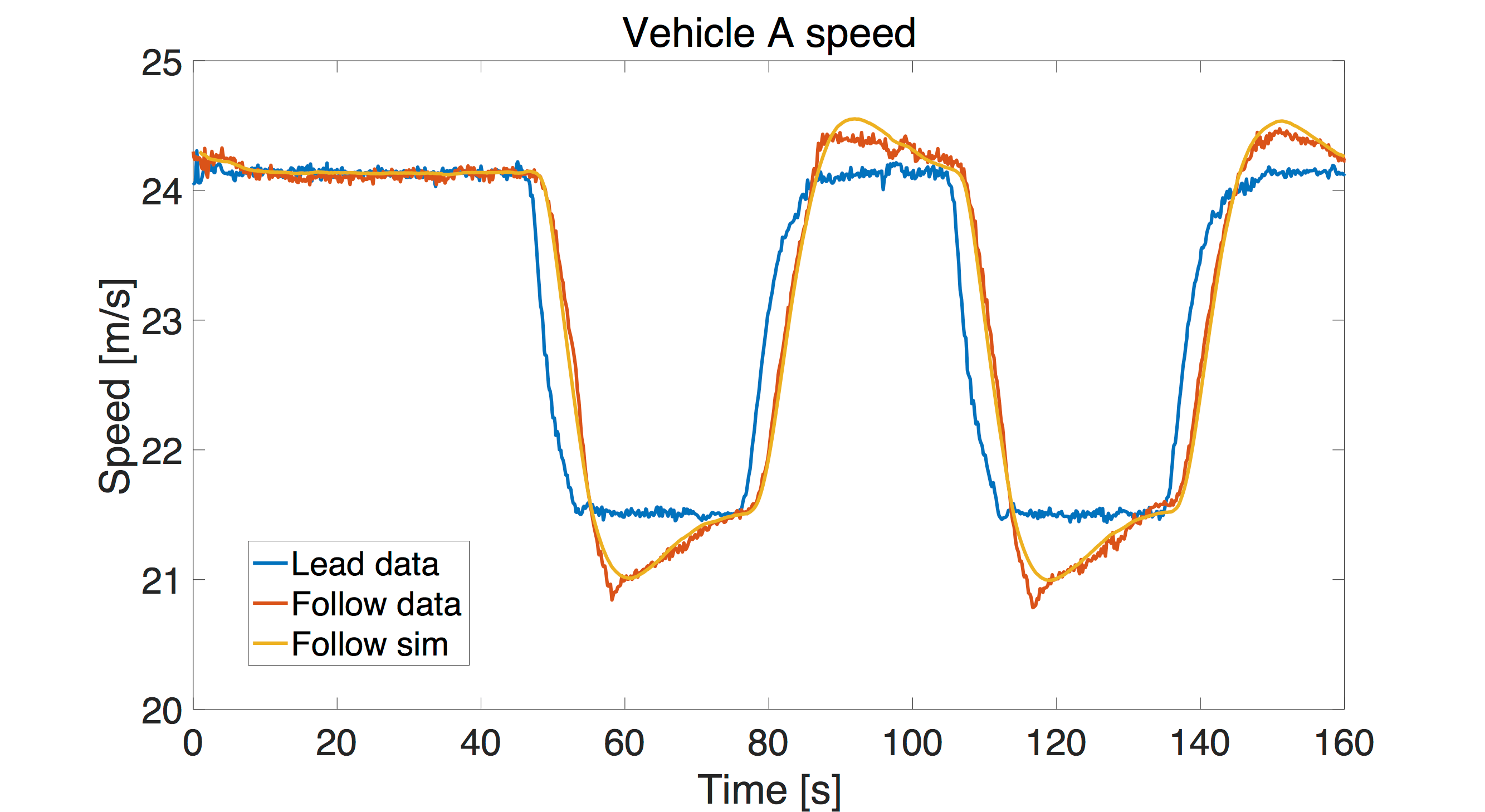}
    \caption{Comparison of calibrated ACC model for the minimum following setting on Vehicle~A between speed and collected experimental data for lead vehicle and follower vehicle on the training data.}
    \label{fig:vehG_test_speed}
\end{figure}

\begin{figure}
    \centering
    \includegraphics[trim=80 0 80 0,clip,width = 0.7\columnwidth]{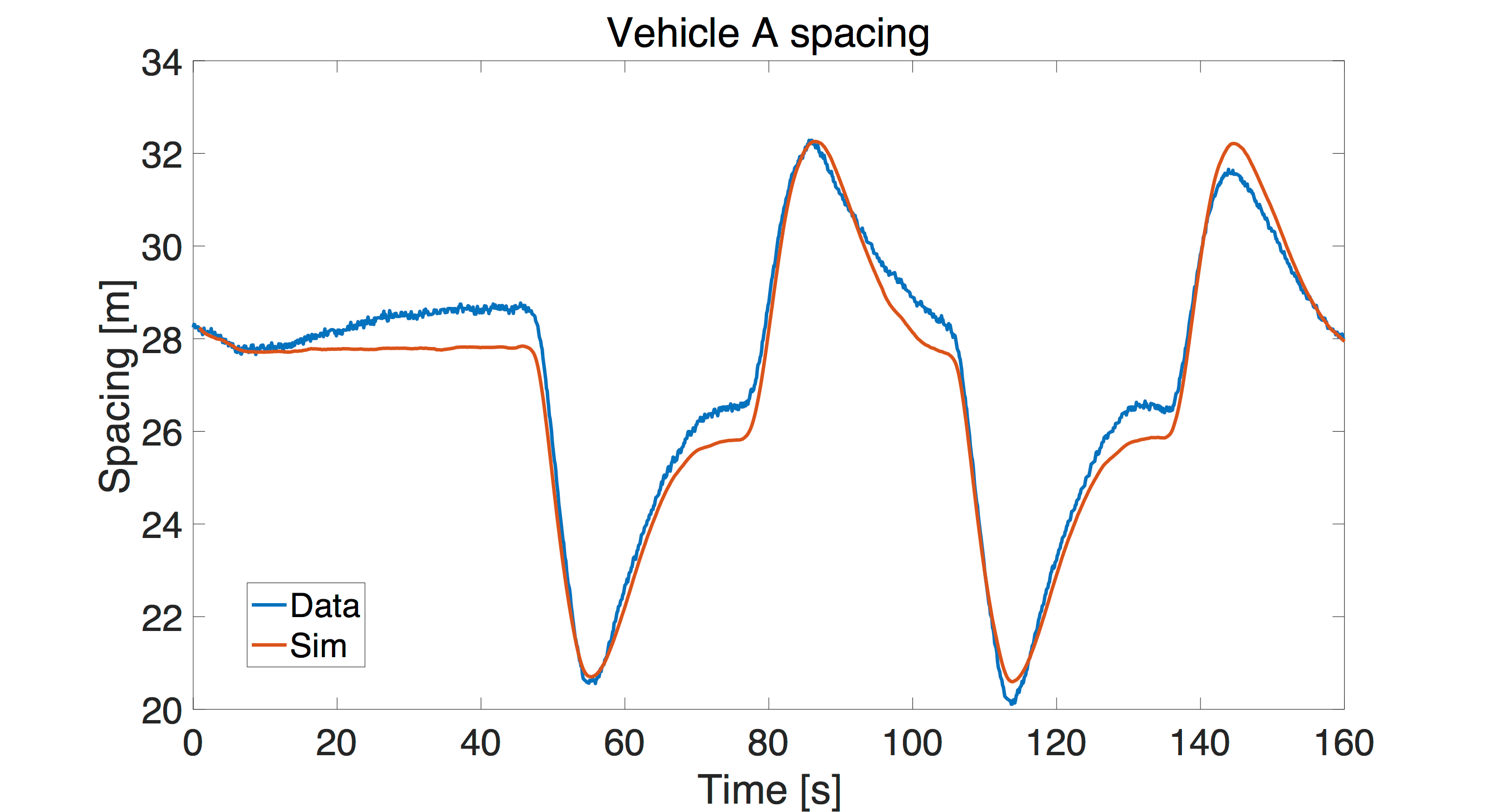}
    \caption{Comparison of calibrated ACC model for Vehicle~A space gap and measured space gap in the training data.}
    \label{fig:vehG_test_space-gap}
\end{figure}

The optimal parameter values for each model are presented in Table~\ref{tab:calibration_stability_results}. An example of the quality of fit of one of the calibrated models is presented in Figure~\ref{fig:vehG_test_speed} where the speed of the lead vehicle along with the recorded and simulated speed of the follower vehicle are plotted, and Figure~\ref{fig:vehG_test_space-gap} where the recorded and simulated space gap are plotted for Vehicle~A at the minimum following setting. Note that the simulated model is able to capture the speed overshoot and undershoot that the follower vehicle exhibits compared to the lead vehicle's speed profile. Similar quality of fit is seen in the spacing plot where the simulated spacing matches the recorded spacing very closely. The plots show performance on training data yet a similar fit is found for the test error (see Table~\ref{tab:test_errors}). Note that we present the results for Vehicle~A, which has the best performance. This is because it is being used in the platoon experiments; and the test error of Vehicle~A for the minimum setting is in line with test errors observed for the remaining vehicles.

The model parameter $k_1$ is the gain for the optimal velocity component and the values range from 0.012~1/s$^2$ to 0.071~1/s$^2$. The model parameter $k_2$ is the gain on the relative velocity component and the calibrated values range from 0.110~1/s to 0.338~1/s indicating a range of behaviors.  The effective time gap $t_h$ of the models range between 0.544~s and 2.054~s. For all vehicles, the value of $t_h$ for the minimum following setting is less than the value of $t_h$ for the maximum following setting, which is expected. Similarly, the values for the sensor lag $\tau$ are all reasonable, and range from 0.153~s for the minimum following setting for Vehicle~B to 0.994~s for the maximum following setting for Vehicle~G. Note that these values include any perception time that the sensors have to compensate for noisy measurements (e.g., through filtering). Finally, the values for the jam space gap $\eta$ range from 5.131 m for the maximum setting for Vehicle~G to 14.956~m for the maximum setting for Vehicle~D. Note here that since Vehicles~A through~D disengage their ACC at speeds below 25 mph, these values of inter-vehicle spacing are not physically attainable. Instead, for these vehicles, these values of inter-vehicle spacing should be thought of as the theoretical jam space gap for this model.

The test error for each model on each collected data set is presented in Table~\ref{tab:test_errors}. The results show that the speed error in simulation is generally quite low with errors as low as 0.100 m/s for Vehicle~A. Note that generally the models perform worst on the data collected during the speed dips tests. The spacing errors in Table~\ref{tab:test_errors} are notably larger, especially as vehicle speeds increase. This is not surprising since at higher speeds, vehicles follow at a greater spacing. Overall, the best fitting model is the model for Vehicle~E for the minimum following setting, which as the lowest speed and spacing training error, while the worst fitting model is the one for Vehicle~B which incurs the highest training error for both speed and spacing. The model with the overall worst performance on any individual test is the one for Vehicle~C for the minimum following setting which has a spacing test error of 9.532 m on the speed dip test, corresponding to a \textit{mean absolute percent error} (MAPE) of 14.01\%. This is because Vehicle~C is a hybrid vehicle with excellent deceleration characteristics but with very modest acceleration. The lack of symmetry between the acceleration behavior of the vehicle and its deceleration behavior with ACC engaged makes it more challenging to fit with the model~\eqref{eq:acc_model}. It was also observed that Vehicle~C takes a long time to close large gaps, which occurs in the speed dip test.

The string stability of each vehicle under the best-fit calibrated parameter values in Table~\ref{tab:calibration_stability_results} is determined using the string stability analysis in Section~\ref{sec:stability}. The result indicates that all vehicles tested are string unstable under both the minimum and the maximum following setting.

\subsection{Platoon simulations under calibrated ACC models}
In this section, the calibrated model for each vehicle is used in simulation for a platoon of vehicles to illustrate the variability of the behavior of the different string unstable ACC systems.
\begin{figure}
    \centering
    \includegraphics[trim=80 0 80 0,clip,width = 0.7\columnwidth]{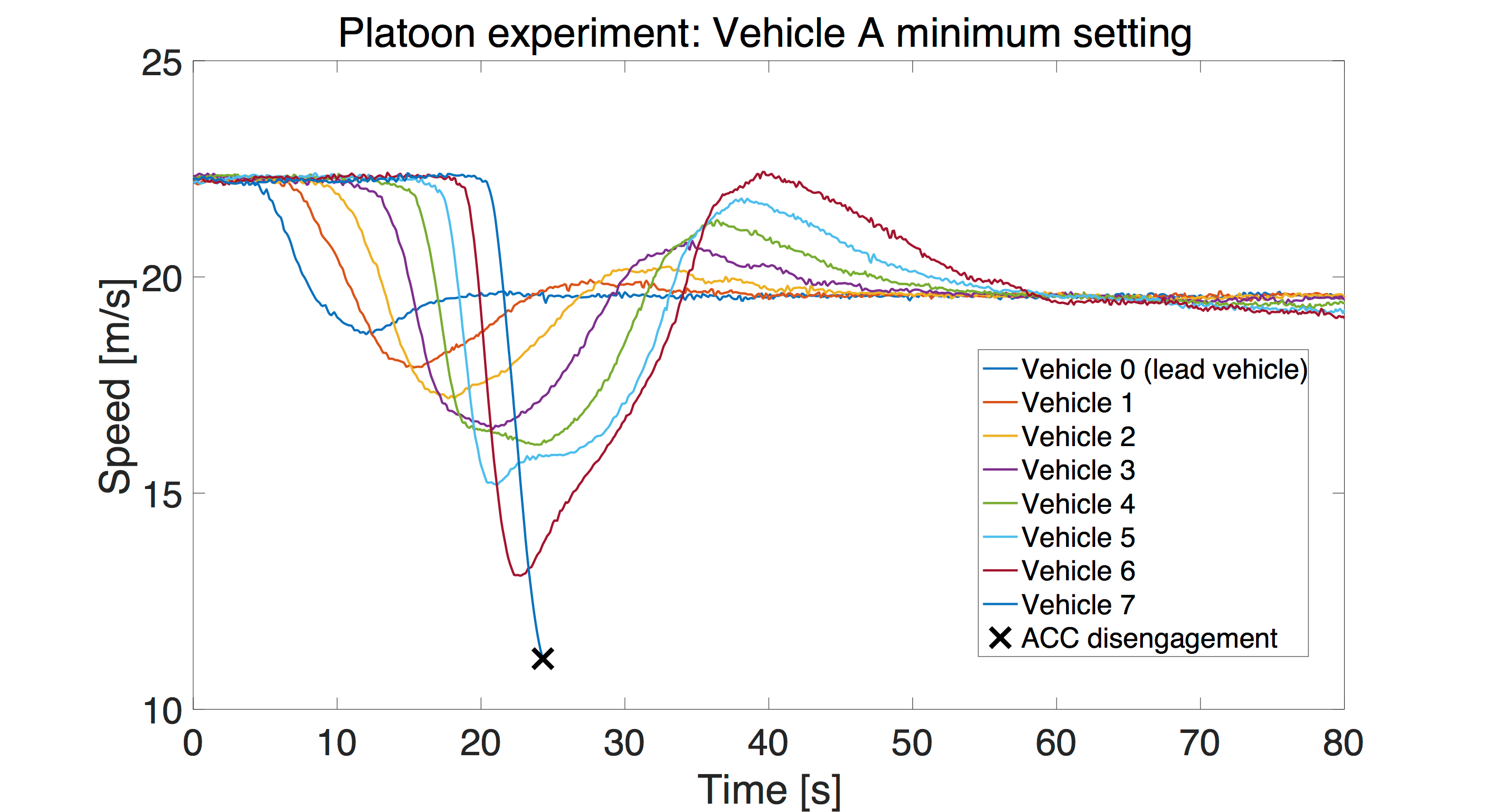}
    \caption{Speeds of vehicles in 8-vehicle platoon test with ACC engaged for last 7 vehicles. Note that due to the instability of the platoon the last vehicle in the platoon slowed to below the minimum operating speed for ACC on the vehicle and switched to driving under human control from that point forward.}
    \label{fig:Platoon_Figure}
\end{figure}

\begin{figure}
    \centering
    \includegraphics[trim=80 0 80 0,clip,width =0.7 \columnwidth]{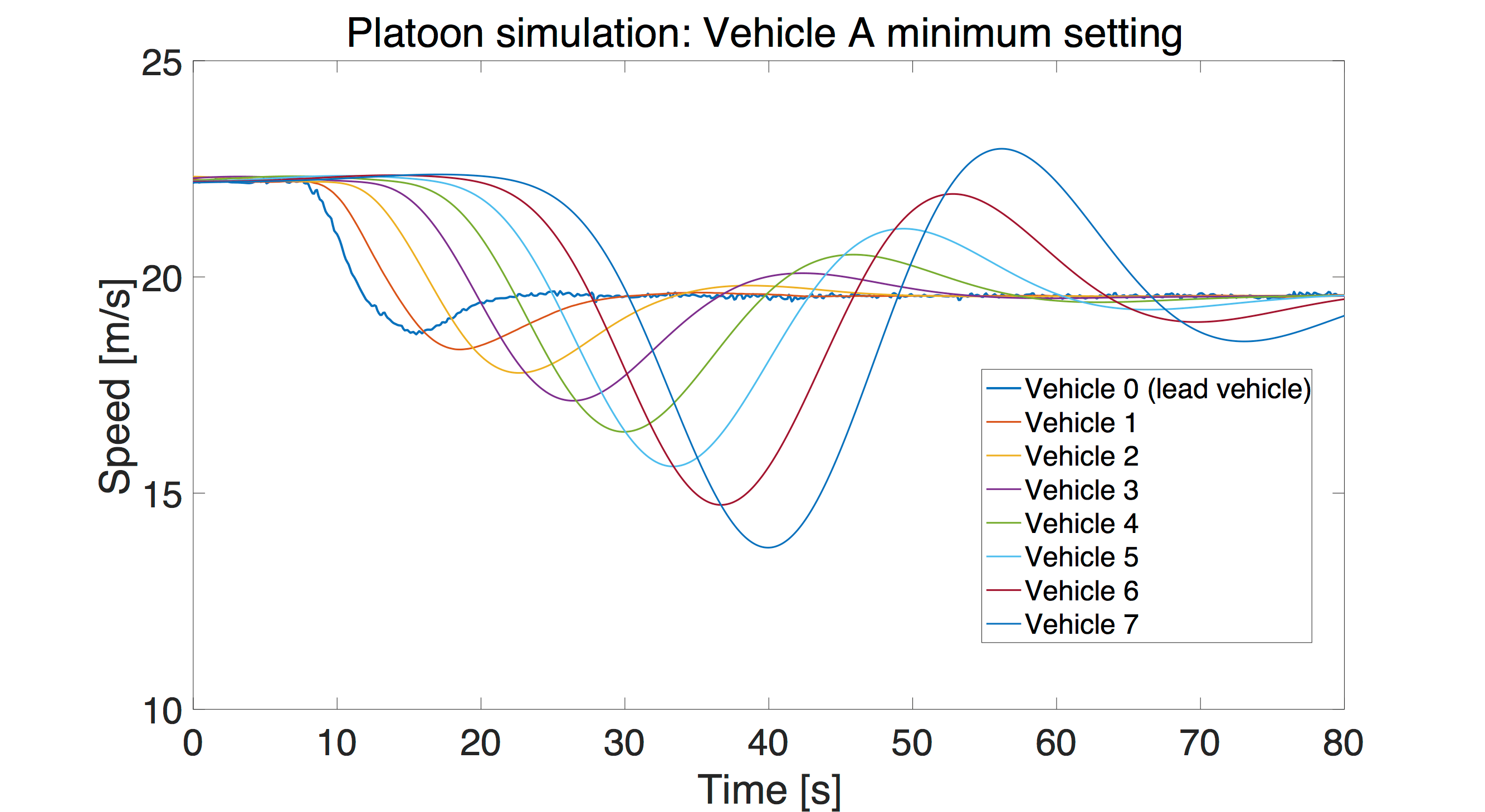}
    \caption{A simulation of an eight vehicle platoon of vehicle A, using the model found for that vehicle in the two-vehicle testing routine.}
    \label{fig:Platoon_Figure_Simulation}
\end{figure}

The result of an eight vehicle platoon (one lead vehicle and seven follower vehicles) simulation are presented in Figure~\ref{fig:Platoon_Figure_Simulation}. The lead vehicle trajectory is taken from the data collected in the platoon experiment described in Section~\ref{sec:platoon_exp}, and the follower vehicles are all simulated using the model found for Vehicle~A at the minimum following setting. The result in in simulation shows that the initial disturbance of 6 mph (2.7 m/s) is amplified by roughly 13 mph (5.8 m/s) to 21 mph (9.4 m/s).

\begin{figure}
    \centering
    \includegraphics[trim=80 0 80 0,clip,width = 0.7\columnwidth]{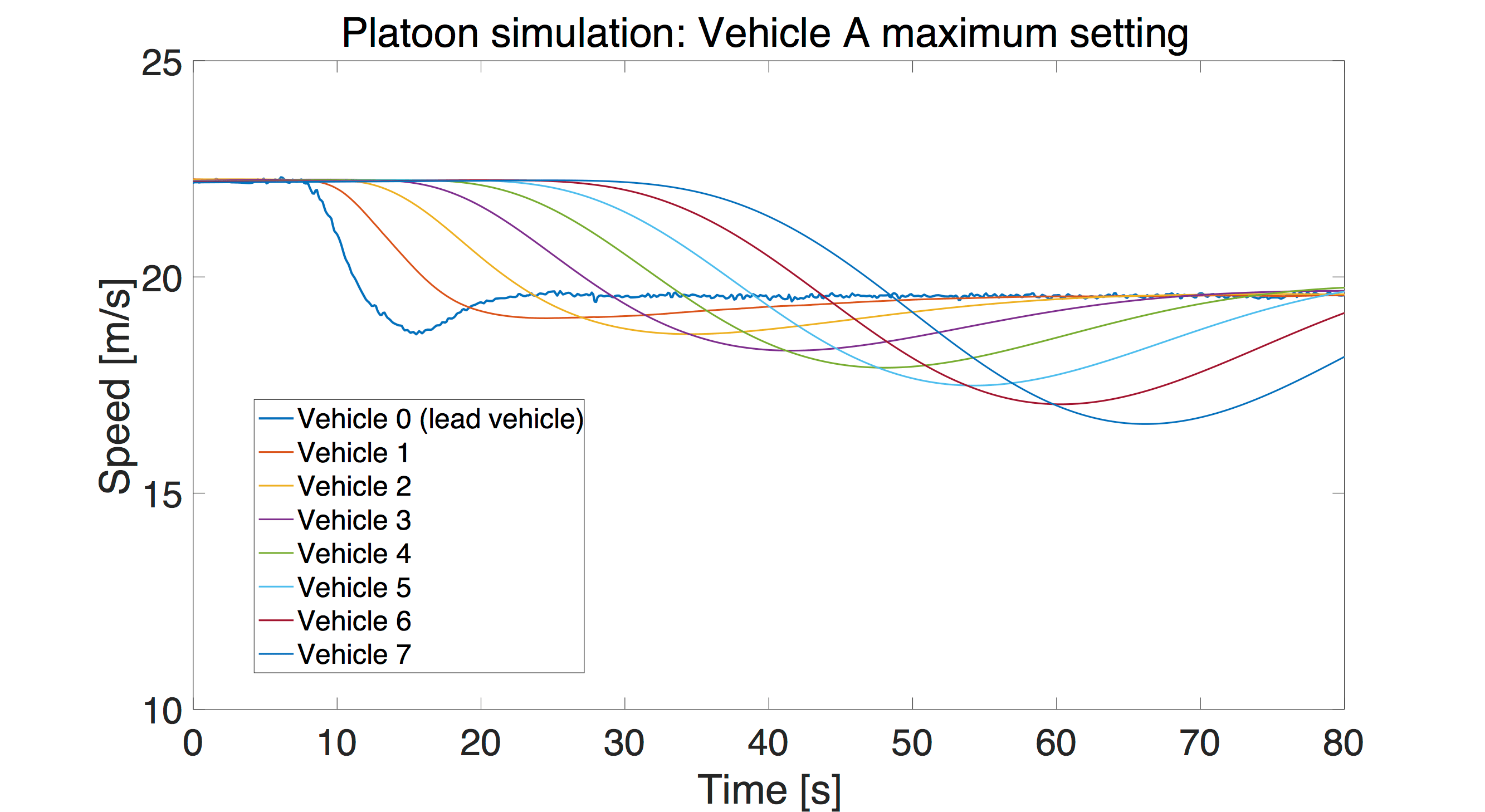}
    \caption{Simulated speed for platoon of eight vehicles using the calibrated model for Vehicle~A with the maximum following setting.}
    \label{fig:max_platoon}
\end{figure}

The same lead vehicle speed profile is used to simulate a platoon with seven follower vehicles using the model for vehicle A under the maximum following setting (Figure~\ref{fig:max_platoon}). The resulting platoon of vehicles is also string unstable, and the initial perturbation amplifies as it propagates from one vehicle to the next. However, the growth rate of the perturbation is much smaller than for the minimum following setting.

\begin{figure}
    \centering
    \includegraphics[trim=0 0 0 0,clip, width = 0.48\columnwidth]{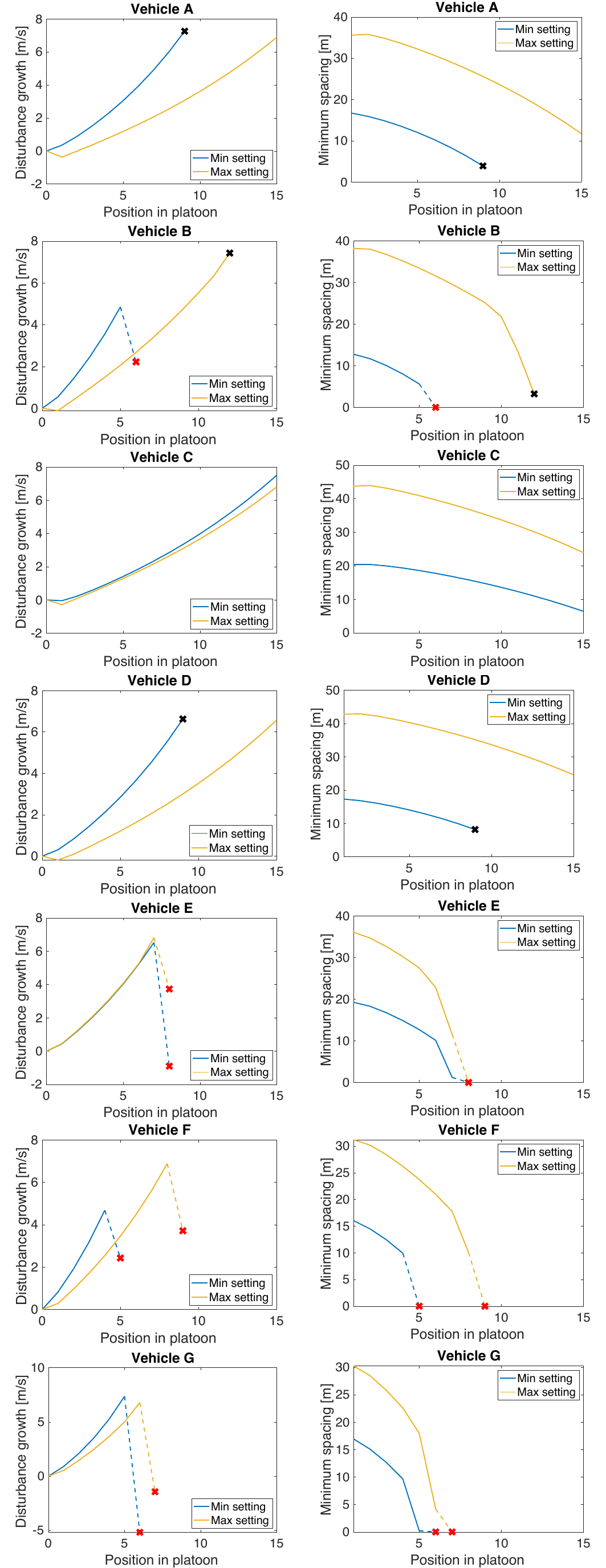}
    \caption{Minimum space gap between vehicles and disturbance amplification for each vehicle and following setting simulated as a function of position in the platoon. A red {\fontfamily{cmss}\selectfont \textbf{x}} indicates that the disengagement occurred because the space gap between two vehicles was zero, while a black {\fontfamily{cmss}\selectfont \textbf{x}} indicates that the disengagement occurred when a vehicle went below the minimum ACC operating speed. Note that the disengagement is plotted at the actual speed or spacing at which this occurs. A dashed line is used to indicate the last vehicle in the platoon, which causes the disengagement.}
    \label{fig:disturbance_growth}
\end{figure}

With the different disturbance amplification behaviors exhibited in the platoon simulations in Figures~\ref{fig:Platoon_Figure_Simulation} and~\ref{fig:max_platoon} in mind, the disturbance amplification and minimum space gap for each vehicle tested as a function of the position in the platoon is plotted in Figure~\ref{fig:disturbance_growth}. This provides insight into the range of disturbance amplification behavior exhibited by the vehicles tested. This plot is constructed by simulating platoons of different length using the calibrated model parameters in Table~\ref{tab:calibration_stability_results} and the lead vehicle speed profile from the lead vehicle in the platoon experiment. For each vehicle, the minimum inter-vehicle spacing during the simulation is plotted as a function of the position in the platoon, and the amplification of the initial 6 mph (2.7 m/s) disturbance is plotted on the right. In cases where the initial disturbance causes vehicles to go below the minimum ACC operating speed given in Table~\ref{tab:vehicle_summary} or the inter-vehicle spacing in simulation becomes negative, the simulation is terminated and an {\fontfamily{cmss}\selectfont \textbf{x}} is plotted for the maximum length platoon for which these constraints are not violated. Specifically, a black {\fontfamily{cmss}\selectfont \textbf{x}} is used when the vehicle speed in simulation goes below the minimum ACC operating speed (i.e., the ACC disengages), while a red {\fontfamily{cmss}\selectfont \textbf{x}} is used when the inter-vehicle space gap goes below zero (i.e., a collision\footnote{We caution the reader that a collision in simulation does not necessarily imply a collision will occur with real vehicles. Collision avoidance systems such as emergency brake assist are not modeled in the simulations presented in this work but are present on real vehicles.} occurs).  These are plotted at the exact speed or spacing for which the disengagement occurs. Therefore, in the case of a collision (i.e., zero inter-vehicle spacing), the speed of the platoon vehicle for which this occurs is faster than the vehicle in front if it, which is indicated with a dashed line.  

The resulting figure shows that the initial disturbance of 6 mph (2.7 m/s) amplifies at different rates for each vehicle and each following setting. For almost all vehicles, the disturbance amplifies faster under the minimum following setting than under the maximum following setting. However, notably, this is not true for Vehicle~E, where the disturbances amplify at almost exactly the same rate. Vehicle~C also exhibits similar behavior to Vehicle~E with the disturbance amplifying at almost the same rate. This may be since Vehicle~C is a hybrid vehicle. Overall, Vehicle~F at the minimum following setting experiences a platoon disengagement for the shortest length platoon with disengagements occurring platoons of length greater than four vehicles, while Vehicle~C does not experience a disengagement for any platoon of length up to 15 vehicles. For longer vehicle platoons, all vehicle models are observed to disengage.

\subsection{Validation via a platoon experiment}
Finally, we discuss the results from the actual eight-vehicle platoon experiment.
In Figure~\ref{fig:Platoon_Figure} the results from the platoon experiment are presented, where a lead vehicle performs a 2.7 m/s (6 mph) slow down with 7 follower vehicles of type~A with ACC engaged on following setting 1. In fact, each vehicle in the platoon exhibits a progressively more extreme braking response than the vehicle before it, which is consistent with the notion that the vehicles are string unstable. Here, for the last follower vehicle the response is large enough in magnitude that the vehicle drops below the minimum speed threshold at which the ACC system is permitted to operate (11.2 m/s, or 25 mph), and control of the vehicle was returned to the human driver.

Comparing the results in the platoon experiment with the simulation results in Figure~\ref{fig:Platoon_Figure_Simulation}, we see that in the experiment, the final vehicle's speed dropped below 11.2 m/s (25 mph), but in simulation only to slightly below 15 m/s (33.6 mph), which would not be enough to cause the 7th follower vehicle in the platoon to have an ACC disengagement. This result, that the simulation is conservative in its prediction of the degree of string instability in the vehicle compared to a real platoon, suggests that the calibrated models may in fact underestimate the extent to which the ACC vehicles tested exhibit string instability in their control and dynamic response.

\section{Conclusion}
This work tested the string stability of adaptive cruise control systems on seven vehicle models from two makes. Using car following data collected on more than 1,200 miles of driving, delay differential equation models of the vehicle under ACC control were calibrated under minimum and maximum following settings. All vehicles under all following settings were found to be string unstable. An eight vehicle platoon test using all identical vehicles confirmed the string instability finding by amplifying an initial 6 mph disturbance by an additional 19 mph, at which point the last vehicle in the platoon dropped below the minimum speed at which the ACC system is operational, and control was handed back to the human driver. 

While string stable ACC system designs have been proposed~(e.g., \cite{rajamani2011vehicle,liang1999optimal}), our emphasis here is in the assessment of the commercial systems now available on many commercial cars as a standard feature. Given that they represent an automation system that has the potential to impact overall traffic flow stability and the occurrence of phantom jams, it is important to not only quantify the system stability but to also provide models that highlight the wide performance variation in the systems. 

Moving forward, higher fidelity ACC models may need to be developed for some vehicle classes that are characterized by distinct acceleration and deceleration behaviors, such as hybrid vehicles. We also caution the reader that commercial ACC systems still have the potential to outperform human drivers with respect to the growth rate of perturbations, which if true would result in a net benefit if such systems are introduced in the flow and operate at a full range of driving speeds. Comparisons of these commercial ACC systems to human drivers in the spirit of the experimental systems tested in~\cite{bose2003analysis} is left for for future work.

\section*{Acknowledgment}
This material is based upon work supported by the National Science Foundation under Grant No. CNS-1446715 (B.P.), CNS-1446690 (B.S.), CNS-1446435 (R.L.), and CNS-1446702 (D.W.).
The authors would like to acknowledge Caroline Janssen, William Barbour, Yanbing Wang, Nancy Emptage, Shelly Wolf, Mary Margaret Sprinkle, Chuck Nicholas and Bozhi Liu for their assistance in organizing and conducting the experiments. The authors would also like to thank the Enterprise representatives at Phoenix and Tucson Airports.

\bibliographystyle{unsrt}
\bibliography{refs_arxiv}

\begin{IEEEbiography}[{\includegraphics[width=1in,height=1.25in,clip,keepaspectratio]{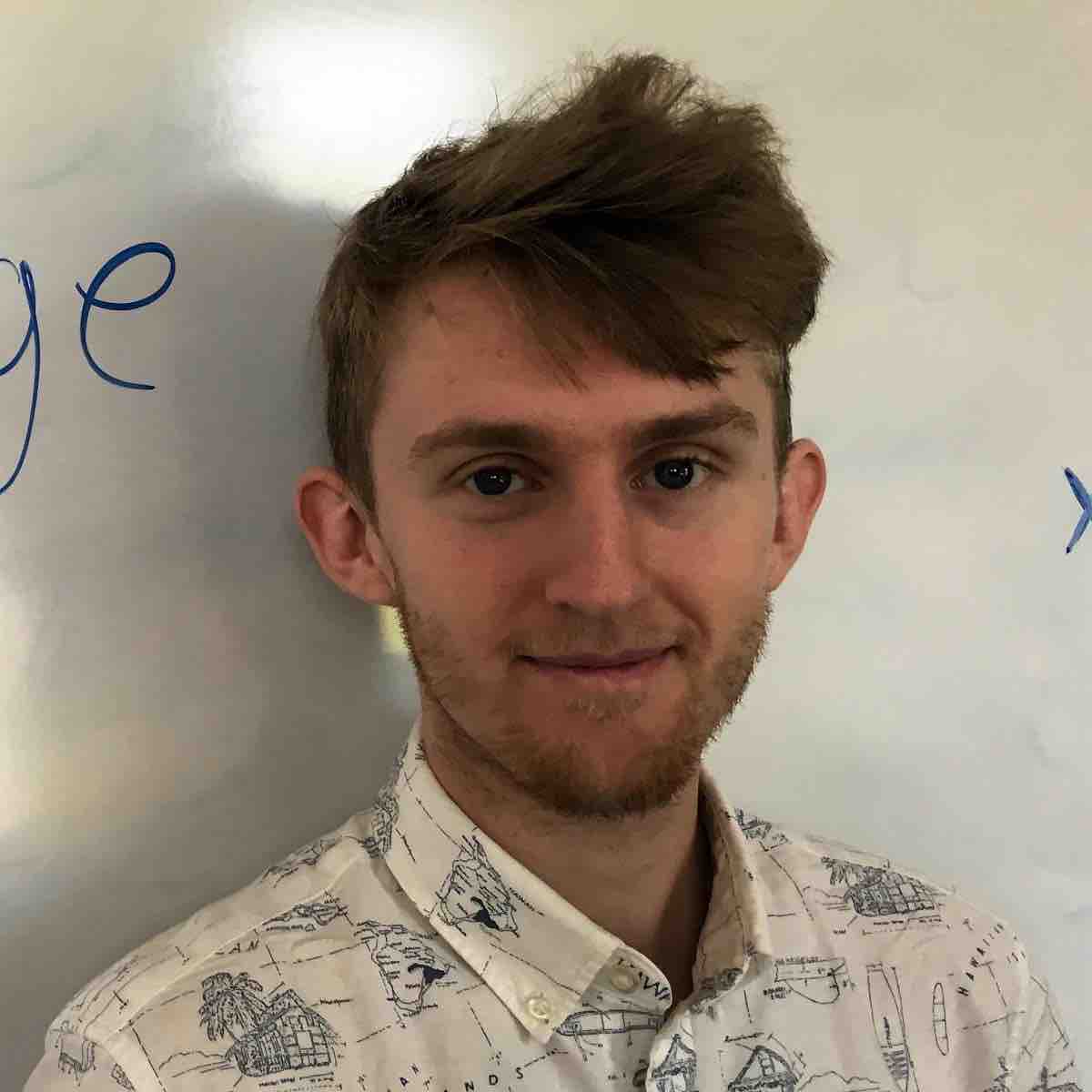}}]{George Gunter} is an undergraduate in the Department of Civil and Environmental Engineering at the University of Illinois. Mr. Gunter is also a visiting undergraduate researcher at the Institute for Software Integrated Systems at Vanderbilt University. His research interests include transportation cyber-physical systems and autonomous vehicles.
\end{IEEEbiography}

\begin{IEEEbiography}[{\includegraphics[width=1in,height=1.25in,clip,keepaspectratio]{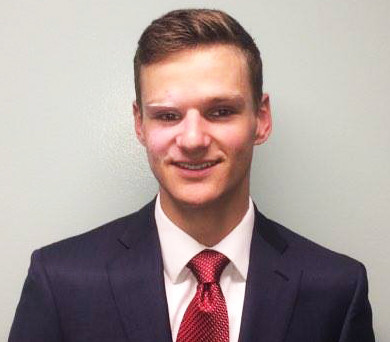}}]{Derek Gloudemans} is a graduate student in the Department of Electrical Engineering and Computer Science and the Institute for Software Integrated Systems at Vanderbilt University. Mr. Gloudemans received his Bachelors Degree in Civil and Environmental Engineering at Vanderbilt University (2018). His research interests include smart cities and intelligent transportation systems.
\end{IEEEbiography}

\begin{IEEEbiography}[{\includegraphics[width=1in,height=1.25in,clip,keepaspectratio]{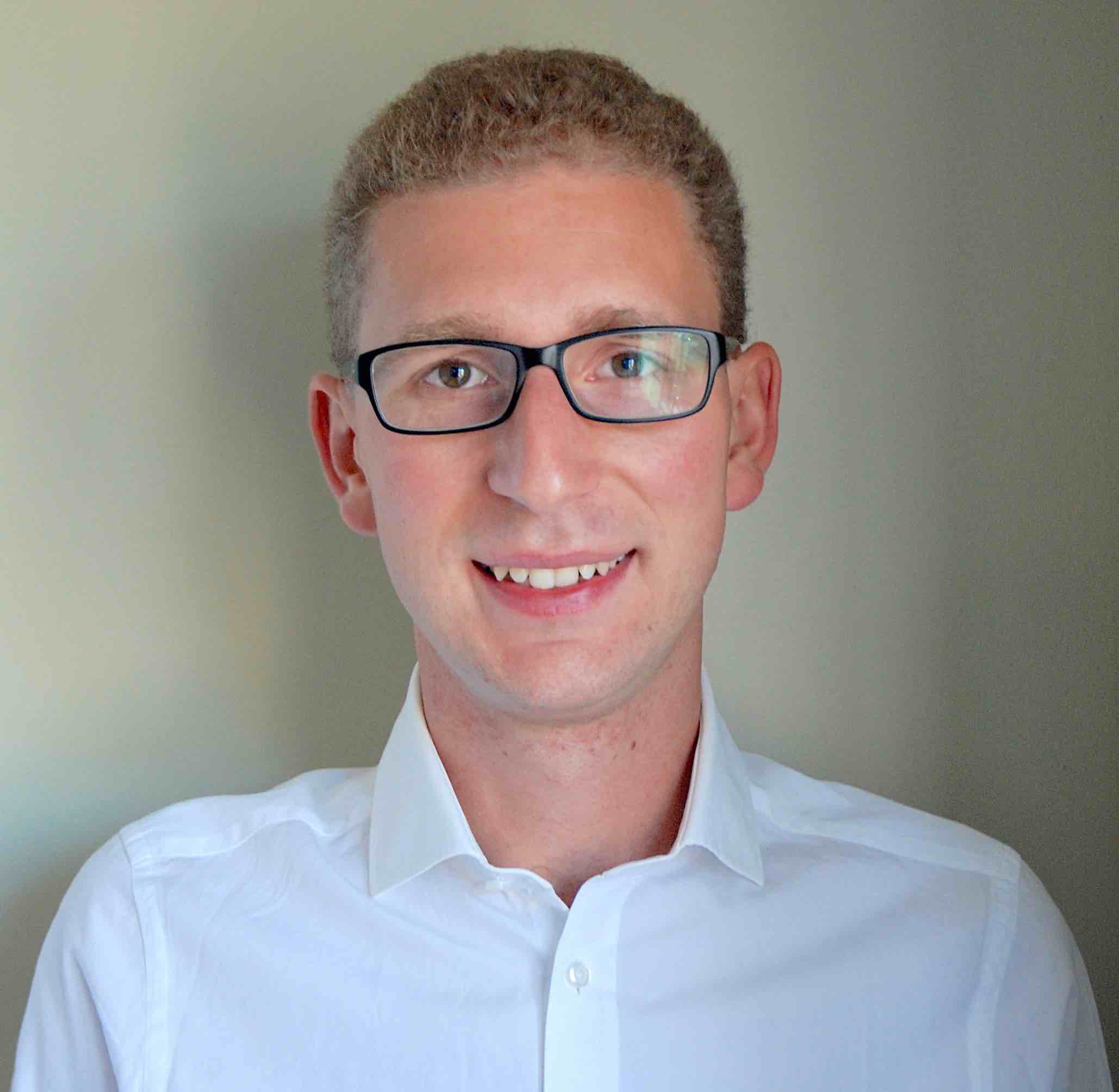}}]{Raphael Stern}
is a visiting researcher at the Institute for Software Integrated Systems at Vanderbilt University. Dr. Stern received a bachelor of science degree (2013), master of science degree (2015), and Ph.D. (2018) all in Civil Engineering from the University of Illinois at Urbana-Champaign. Dr. Stern was a visiting researcher at the Institute for Pure and Applied Mathematics at UCLA, and a recipient of the Dwight David Eisenhower Graduate Fellowship from the Federal Highway Administration. Dr. Stern's research interests are in the area of traffic control and estimation with autonomous vehicles in the flow.
\end{IEEEbiography}

\begin{IEEEbiography}[{\includegraphics[width=1in,height=1.25in,clip,keepaspectratio]{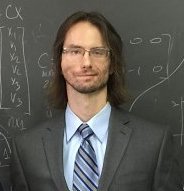}}]{Sean McQuade} is a PhD student in Applied Mathematics at Rutgers University -- Camden. Prior to joining Rutgers University, he received a MS in Risk Management from Temple University and a BS in Mathematics from Virginia Polytechnic Institute and State University. His research interests include mathematical models for transportation systems and metabolic networks. He is a recipient of a 2018 research fellowship from the Mistletoe Foundation. 
\end{IEEEbiography}

\begin{IEEEbiography}[{\includegraphics[width=1in,height=1.25in,clip,keepaspectratio]{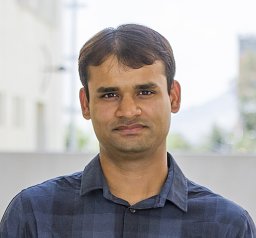}}]{Rahul Bhadani} is a PhD student in the Department of Electrical and Computer Engineering at the University of Arizona. Mr. Bhadani holds a BS degree in Information Technology with an emphasis on Computer Science and Software Engineering. His research interests include modeling, simulation and control of autonomous vehicles, developing novel statistical models for traffic simulation. Prior to joining the University of Arizona, Mr. Bhadani worked as software engineer for Oracle.
\end{IEEEbiography}

\begin{IEEEbiography}[{\includegraphics[width=1in,height=1.25in,clip,keepaspectratio]{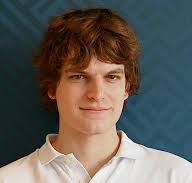}}]{Matt Bunting} is a PhD student in the Department of Electrical and Computer Engineering at the University of Arizona. Prior to his graduate studies, Mr. Bunting received his BS in Electrical and Computer Engineering from the University of Arizona.
\end{IEEEbiography}

\begin{IEEEbiography}[{\includegraphics[width=1in,height=1.25in,clip,keepaspectratio]{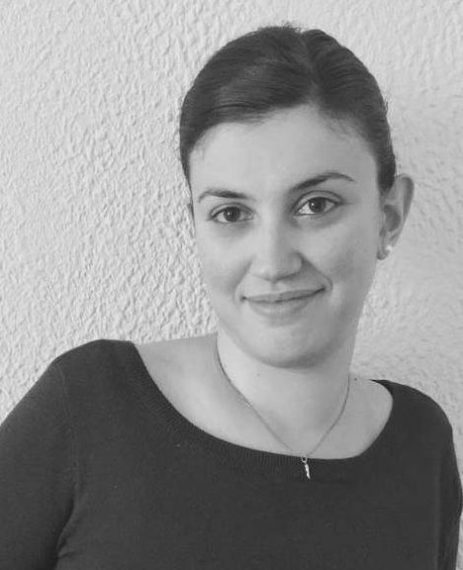}}]{Maria Laura Delle Monache} is a research scientist in the Networked Controlled Systems team at Inria  and in GIPSA-Lab  (Department of Control) in Grenoble. Her research interest is mainly related to the mathematical and engineering aspects of traffic flow. In particular, she is interested in mathematical modeling, analysis, numerical approximation and control of traffic flow applications. Prior to Inria, she was a Postdoctoral researcher at Rutgers University Camden.
\end{IEEEbiography}

\begin{IEEEbiography}[{\includegraphics[width=1in,height=1.25in,clip,keepaspectratio]{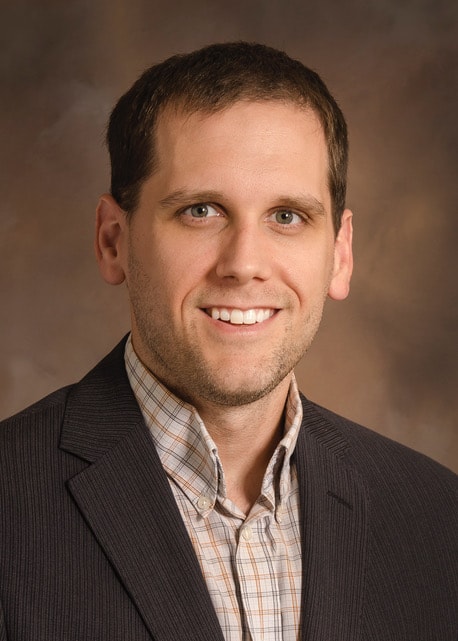}}]
{Roman Lysecky} is a Professor of Electrical and Computer Engineering at the University of Arizona. He received his Ph.D. in Computer Science from the University of California, Riverside in 2005. His research focuses on embedded systems with emphasis on medical device security, automated threat detection and mitigation, runtime adaptable systems, performance and energy optimization, and non-intrusive observation methods. He is an author on more than 100 research publications in top journals and conferences. He received the Outstanding Ph.D. Dissertation Award from the European Design and Automation Association (EDAA) in 2006, a CAREER award from the National Science Foundation in 2009, and seven Best Paper Awards.
\end{IEEEbiography}

\begin{IEEEbiography}[{\includegraphics[width=1in,height=1.25in,clip,keepaspectratio]{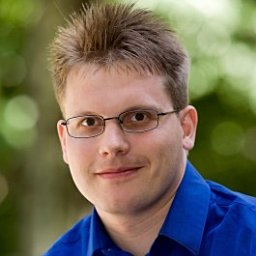}}]{Benjamin Seibold} is the Director of the Center for Computational Mathematics and Modeling and an associate professor of Mathematics at Temple University. He received his Dr.rer.nat. (2006) from the University of Kaiserslautern, Germany, and he was an Instructor of Applied Mathematics at the Massachusetts Institute of Technology. His research expertise includes high-order methods for fluid flows and interface evolution, radiative transfer and kinetic problems, and traffic flow modeling, simulation, and control.
\end{IEEEbiography}

\begin{IEEEbiography}[{\includegraphics[width=1in,height=1.25in,clip,keepaspectratio]{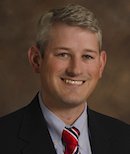}}]{Jonathan Sprinkle} (S'96-M'03-SM'11) received the B.S. degree from Tennessee Technological University, Cookeville, TN, USA, in 1999, and the M.S. and Ph.D. degrees from Vanderbilt University, Nashville, TN, USA, in 2000 and 2003, respectively. He is the Litton Industries John M. Leonis Distinguished Associate Professor of Electrical and Computer Engineering, University of Arizona, Tucson, AZ, USA. From 2003 to 2007, he was at the University of California, Berkeley, CA, USA, as a Postdoctoral Scholar. He joined the University of Arizona in 2007. His research interests and experience are in systems control and engineering, and he teaches courses ranging from systems modeling and control to mobile application development and software engineering. Prof. Sprinkle received the NSF CAREER Award in 2013, and in 2009, he received the UA's Ed and Joan Biggers Faculty Support Grant for work in autonomous systems. His work has an emphasis for industry impact, and he was recognized with the UA ``Catapult Award'' by Tech Launch Arizona in 2014, and in 2012, his team won the NSF I-Corps Best Team Award.
\end{IEEEbiography}

\begin{IEEEbiography}[{\includegraphics[width=1in,height=1.25in,clip,keepaspectratio]{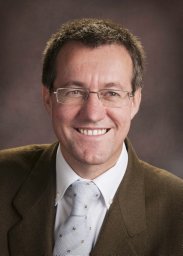}}]{Benedetto Piccoli} received the Ph.D. degree in applied mathematics from the Scuola Internazionale Superiore di Studi Avanzati (SISSA), Trieste, Italy, in 1994. He was a Researcher with the SISSA from 1994 to 1998, an Associate Professor with the University of Salerno from 1998 to 2001, and a Research Director with Istituto per le Applicazioni del Calcolo ``Mauro Picone'' of the Italian Consiglio Nazionale delle Ricerche (IAC-CNR), Rome, Italy, from 2001 to 2009. Since 2009, he has been the Joseph and Loretta Lopez Chair Professor of Mathematics with the Department of Mathematical Sciences, Rutgers University, Camden, NJ, USA.
\end{IEEEbiography}

\begin{IEEEbiography}[{\includegraphics[width=1in,height=1.25in,clip,keepaspectratio]{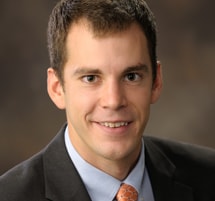}}]{Daniel B. Work}
is an associate professor in Civil and Environmental Engineering and Institute for Software Integrated Systems at Vanderbilt University. Prof. Work earned his B.S. degree (2006) from the Ohio State University, and an M.S. (2007) and Ph.D. (2010) from the University of California, Berkeley, each in civil engineering. His research interests include transportation cyber physical systems.  He is a recipient of the CAREER award from the National Science Foundation (2014) the Gilbreth Lectureship from the National Academy of Engineering (2018).
\end{IEEEbiography}

\end{document}